\documentclass[pra,twocolumn,floatfix,aps]{revtex4}
\usepackage{amsmath}
\usepackage{makeidx}
\usepackage{amssymb}
\usepackage{amsmath}
\usepackage{graphicx}

\usepackage[usenames]{color}
\usepackage[normalem]{ulem}
\usepackage{gensymb}

\usepackage{hyperref}
\usepackage[T1]{fontenc} 
\begin{document}

\title{Supersolid-like square- and honeycomb-lattice crystallization of droplets  in a dipolar 
condensate}

\author{Luis E. Young-S.$^1$\footnote{lyoung@unicartagena.edu.co}}

 \author{S. K. Adhikari$^2$\footnote{sk.adhikari@unesp.br,  professores.ift.unesp.br/sk.adhikari/}}

 \affiliation{$^1$Grupo de Modelado Computacional, Facultad de Ciencias Exactas y Naturales, Universidad de Cartagena, 
130014 Cartagena, Bolivar, Colombia}

\affiliation{$^2$Instituto de F\'{\i}sica Te\'orica, UNESP - Universidade Estadual Paulista, 01.140-070 S\~ao Paulo, S\~ao Paulo, Brazil}

\begin{abstract}

We demonstrate a supersolid-like spatially-periodic square- and honeycomb-lattice  crystallization of droplets, in addition to the commonly-studied triangular-lattice crystallization, 
in a cylindrically-symmetric quasi-two-dimensional trapped dipolar condensate, using a beyond-mean-field model including a quantum-fluctuation Lee-Huang-Yang-type
 interaction.  These three types of crystallization  of droplets may appear for the {\it same} atomic interactions and the {\it same} trap frequencies.
 The energy $E$ of all three  crystallization  as a function of
 number $N$ of atoms satisfy the universal scaling relation $E\sim N^{0.4}$ 
  indicating that all  three arrangements of the droplets should be energetically probable processes  of phenomenological interest.   The state of square-lattice crystallization may have the central site  occupied or   unoccupied, corresponding to a parity-symmetric or parity-antisymmetric state, respectively.  The state of square-lattice crystallization with the occupied central site and the state of triangular-lattice crystallization, for a fixed $N$,  constitute two quasi-degenerate ground states while the other states are low-lying excited states.  This makes the square-lattice crystallization with the occupied central site an ideal candidate for future experimental observation.

\end{abstract}


\maketitle

\section{Introduction}

A supersolid \cite{sprsld,sprsld1,sprsld2,sprsld3,sprsld4,sprsld5} is a special form of quantum matter, which exhibits a  spatially-ordered stable structure, as encountered in a 
solid crystal,  breaking continuous translational invariance. A supersolid   can also  flow without friction  as a superfluid breaking  continuous gauge invariance. Hence, contrary to the wisdom that a frictionless flow is an exclusive property of a superfluid, 
a supersolid  simultaneously possesses the properties of a superfluid and a solid. The search
of  supersolidity in   $^4$He \cite{4} was not conclusive \cite{5}. However, there had been theoretical suggestions for creating a supersolid in a  dipolar Bose-Einstein condensate (BEC) \cite{santos,7a,7b}, {in a BEC with finite-range atomic interaction \cite{xxx}}   and in a   spin-orbit (SO) coupled spinor BEC \cite{7c}.
The study of a supersolid has recently gained new impetus among research workers in low-temperature physics, 
after the experimental  observation of supersolids in a dipolar BEC  \cite{2d3,drop1} and in an SO-coupled pseudo spin-1/2 spinor BEC \cite{18}.

Recently,
a spatially-periodic  state, displaying a  stripe   pattern in density, known as  a superstripe state, because of its supersolid-like properties, was experimentally observed
in an SO-coupled pseudo spin-1/2  BEC
of $^{23}$Na atoms \cite{18}. {
In a quasi-two-dimensional (quasi-2D)  uniform 
and trapped    SO-coupled spin-2 \cite{sandeep},  spin-1   \cite{so1-1,so1-3}, and pseudo spin-1/2 \cite{so1/2} spinor BEC,   the formation of  a  {supersolid-like} {
\it square- and triangular-lattice} patterns in density was demonstrated in theoretical studies  in addition to the superstripe state \cite{7c,18}.  }
In a strongly dipolar BEC, for an appropriate mixture of dipolar and contact interactions, and for the number of atoms $N$  beyond a critical value, 
high-density droplet formation was  observed experimentally  in a  dipolar BEC under a strong trap   of $^{164}$Dy \cite{drop1} and  $^{168}$Er \cite{drop2} atoms and 
studied theoretically \cite{drop3,blakie}.  In the framework of a mean-field model  employing the Gross-Pitaevskii (GP) equation,
a dipolar BEC collapses for a strong dipolar interaction beyond a critical value, 
and 
 a Lee-Huang-Yang-type \cite{lhy} (LHY-type)  beyond-mean-field quantum-fluctuation interaction \cite{qf1,qf2}
is necessary   in theoretical studies   to stabilize a strongly dipolar
droplet against collapse \cite{santos}. As the number of atoms $N$ in a trapped  dipolar BEC is increased, so that the density of atoms reaches a critical value, due  to the dipolar interaction, the condensate shrinks to a very small size.  However, it cannot  collapse due to the quantum-fluctuation LHY interaction and a droplet is formed.
 The size of the   droplet is  much smaller than the harmonic oscillator trap lengths.
Such droplets can accommodate a maximum number of atoms \cite{drop3} for  given harmonic trap frequencies so as to attain a critical density of atoms in the condensate. 
  In spite of the name droplet, the present dipolar BEC  droplets in a strong trap  are different from recently observed \cite{binary,binary2}    nondipolar binary  BEC droplets in free space. Nevertheless, in both cases, the collapse is arrested by a  beyond-mean-field quantum-fluctuation LHY interaction.




For a sufficiently large $N$, 
in a quasi-one-dimensional (quasi-1D) trapped dipolar BEC, a spontaneous periodic crystallization of droplets along a straight line 
was observed in different experiments  on $^{164}$Dy \cite{1d1,1d4,1d5},  $^{162}$Dy \cite{1d2,1d3,1d7}, and $^{166}$Er \cite{1d4,1d5}  atoms and confirmed in  related theoretical studies \cite{1d6,1d8}, whereas in a quasi-2D  trapped dipolar BEC of $^{164}$Dy atoms,  a crystallization of droplets on 
a periodic triangular lattice was observed experimentally \cite{2d2,2d3} and  established in theoretical studies \cite{2d4,other1,other2,blakieprl,other3}.  
In addition to this  periodic triangular-lattice state, in a trapped quasi-2D dipolar BEC,
the formation of  honeycomb,  stripe and other
periodic  structures   in density, and not  crystallization of droplets, have also been predicted 
\cite{other1,other3,other2,fau} in theoretical studies. 
 { Nevertheless, in many of these investigations, specially in the numerical studies on a truncated finite system,  the supersolidity of the system has never  been rigorously established \cite{donner}. One needs to show
the spontaneous breaking of gauge symmetry (that gives the superfluid
order parameter) and the spontaneous breaking of translational symmetry in the same system. Lacking a rigorous demonstration of supersolidity, we prefer to call these periodic states supersolid-like states in this paper as in  similar studies on  quasi-2D SO-coupled spinor BECs \cite{so1-3,sandeep}.}

Following the 1D crystallization of dipolar droplets along a straight line in a quasi-1D trap
\cite{1d1,1d2,1d3,1d4,1d5,1d6,1d7,1d8},
the natural crystallization  in 2D is the square-lattice arrangement of droplets  $-$ not yet observed in experiments and not predicted theoretically. 
{In this paper, using a beyond mean-field model  including the  quantum-fluctuation LHY interaction \cite{qf1,qf2} for a three-dimensional (3D) trapped dipolar BEC, we explicitly demonstrate a {supersolid-like spatially-periodic}    square-  and  honeycomb-lattice crystallization  of droplets in the $x-y$ plane, perpendicular to the polarization $z$ direction, for an appropriate mixture of dipolar and contact interactions in a quasi-2D trap
in addition to the triangular-lattice  crystallization of droplets found in different theoretical \cite{2d4,other1} and experimental \cite{2d2,2d3} investigations. }
{ We found that the symmetry of the final state is sensitive to the initial state employed in numerical simulation.  A final state with a specific symmetry $-$ a square, triangular, or a honeycomb lattice $-$ can be easily obtained with the use of  an initial state with the same symmetry. }
{  No such supersolid-like state can be obtained in a trapped BEC with isotropic contact interaction. In case of dipolar interaction, a single droplet can be stable for a maximum number of atoms. As the number of atoms is increased further, multiple droplets are generated and due to an  interplay between the  dipolar repulsion in the $x-y$ plane and the external trapping potential, a supersolid-like arrangement of droplets is formed.} 


{
In this study we find two distinct types of square-lattice arrangements of dipolar droplets in a circularly-symmetric quasi-2D trapped dipolar BEC, e.g.,  with the central site at $x=y=0$ occupied or vacant.
In the case,  the central site is occupied (unoccupied) by a droplet  the wave function is parity-symmetric 
(parity-antisymmetric).  In the first type we find 9, 25, and 49... droplets arranged on $3\times 3$, $5\times 5$, $7\times 7$... arrays, whereas in the second type we find 4, 16, 36... droplets arranged on $2\times 2$, $4\times 4$, $6\times 6$... arrays as in Fig. \ref{fig2}. Like usual parity-antisymmetric states, square-lattice crystallization with a vacant central site  is an excited state. 
We also numerically investigate the triangular-lattice arrangement of droplets studied  previously.  In addition to the triangular and  square-lattice arrangements, we also demonstrate a clean honeycomb-lattice arrangement of droplets. A honeycomb lattice is a special case of a triangular  lattice with missing  droplets at the centers of adjacent hexagons.  Of these different possibilities,  the triangular-lattice arrangement of droplets and the square-lattice arrangement with an occupied central site constitute two quasi-degenerate stable ground states. { The honeycomb-lattice and the square-lattice arrangements with a vacant central site have slightly larger energies and are excited states. }

 We also calculated the energies of the different states and 
  established a universal scaling relation between the energy per atom $E$ of the supersolid-like  crystallization of droplets on 
square, honeycomb and triangular lattices and the number of atoms  $N,$  independent of the type of lattice,  which implies 
that these three periodic crystallization of droplets are all equally probable energetically. Moreover,  the three different crystallization of  droplets appear for the same atomic contact and dipolar interactions and for the same trap frequencies.  Hence all these periodic crystallization  of droplets should be  of experimental interest.
With this in mind, in this paper, we have employed the same confining trap frequencies and similar number of $^{164}$Dy atoms as in previous experimental \cite{2d2}
and theoretical \cite{2d4}
studies  on the triangular lattice formation of dipolar droplets. 
The number of droplets $n_{\mathrm{d}}$ is found to increase approximately linearly with   $N$.
}

In Sec. \ref{II} we present the beyond-mean-field model including the quantum-fluctuation LHY interaction in the GP equation. The time-independent version of this equation
is also obtained  from a variational rule using a time-independent energy functional.
In Sec. \ref{III} we present the numerical results for stationary  states with  three types of periodic array of droplets, e.g. square lattice, triangular lattice and honeycomb lattice, in a trapped  dipolar BEC.  
    Finally, in Sec. \ref{IV} we present a summary of our findings.

\section{Beyond-Mean-field model}

\label{II}

In this paper we base our study on a  3D beyond-mean-field model   including the  quantum-fluctuation LHY interaction. 
We consider a  BEC of $N$ dipolar  atoms polarized along the $z$ axis, of mass $m$ each,
interacting through the following 
atomic dipolar and contact interactions   \cite{dipbec,dip,yuka}
\begin{eqnarray}
V({\bf R})= 
\frac{\mu_0 \mu^2}{4\pi}\frac{1-3\cos^2 \theta}{|{\bf R}|^3}
+\frac{4\pi \hbar^2 a}{m}\delta({\bf R }),
\label{eq.con_dipInter} 
\end{eqnarray}
where $a$ is the scattering length, $\mu_0$ is the permeability of vacuum,  $\mu$ is the magnetic dipole moment of each atom,
${\bf R}= {\bf r} -{\bf r}'$ is the vector joining two dipoles placed at $\bf r \equiv \{x,y,z\}$ and $\bf r' \equiv \{x',y',z'\}$
and $\theta$ is the angle made by  ${\bf R}$ with the  
$z$ axis. The strength of dipolar 
interaction is given by  the  dipolar length 
\begin{align}
a_{\mathrm{dd}}=\frac{\mu_0 \mu^2 m }{ 12\pi \hbar ^2}.
 \label{eq.dl}
 \end{align}
The dimensionless ratio 
 \begin{equation}
\varepsilon_{\mathrm{dd}}\equiv \frac{a_{\mathrm{dd}}}{a}
\end{equation} 
determines
the strength of the dipolar interaction relative to  the contact interaction 
and controls many properties of a dipolar BEC.

A dipolar BEC is described by the following  3D beyond-mean-field GP equation including the  quantum-fluctuation LHY interaction \cite{dipbec,dip,2d4,blakie,yuka}
\begin{align}\label{eq.GP3d}
 i \hbar \frac{\partial \psi({\bf r},t)}{\partial t} &=\
{\Big [}  -\frac{\hbar^2}{2m}\nabla^2
+U({\bf r})
+ \frac{4\pi \hbar^2}{m}{a} N \vert \psi({\bf r},t) \vert^2 \nonumber\\
&\ +\frac{3\hbar^2}{m}a_{\mathrm{dd}}  N
\int \frac{1-3\cos^2 \theta}{|{\bf R}|^3}
\vert\psi({\mathbf r'},t)\vert^2 d{\mathbf r}'  
\nonumber 
 \\
& +\frac{\gamma_{\mathrm{QF}}\hbar^2}{m}
|\psi({\mathbf r},t)|^3
\Big] \psi({\bf r},t), 
\end{align}
where $U({\bf r})=\frac{1}{2}m(\omega_x^2x^2+\omega_y^2y^2+\omega_z ^2z^2) $ is the trap with angular frequencies $\omega_x\equiv 2\pi f_x, \omega_y \equiv 2\pi f_y, \omega_z\equiv 2\pi f_z$ along $x,y,z$ directions, respectively,
the wave function is  normalized as $\int \vert \psi({\bf r},t) \vert^2 d{\bf r}=1$. The coefficient 
of the beyond-mean-field  quantum-fluctuation LHY term $\gamma_{\mathrm{QF}}$ is given by \cite{qf1,qf2,blakie}
\begin{align}\label{qf}
\gamma_{\mathrm{QF}}= \frac{128}{3}\sqrt{\pi a^5} Q_5(\varepsilon_{\mathrm{dd}}),
\end{align}
where the auxiliary function
\begin{equation}
 Q_5(\varepsilon_{\mathrm{dd}})=\ \int_0^1 dx(1-x+3x\varepsilon_{\mathrm{dd}})^{5/2} 
\end{equation}
 can be evaluated as \cite{blakie}
\begin{align}\label{exa}
Q_5(\varepsilon_{\mathrm{dd}}) &=\
\frac{(3\varepsilon_{\mathrm{dd}})^{5/2}}{48}  \mathrm{Re}\left[(8+26\epsilon+33\epsilon^2)\sqrt{1+\epsilon}\right.\nonumber\\
& + \left.
\ 15\epsilon^3 \mathrm{ln} \left( \frac{1+\sqrt{1+\epsilon}}{\sqrt{\epsilon}}\right)  \right], \quad  \epsilon = \frac{1-\varepsilon_{\mathrm{dd}}}{3\varepsilon_{\mathrm{dd}}}, \\
&\approx 1+ \frac{3}{2}\varepsilon_{\mathrm{dd}}^2  \label{app}
\end{align}
where Re denotes the real
 part. {In the present study we use the exact expression (\ref{exa}). {Actually, for $ \varepsilon_{\mathrm{dd}}
>1$,  $Q_5$ is complex and its small imaginary part will be  neglected, as in other studies [37,38], in the present study of stationary droplet states.}
 The use of the approximate expression (\ref{app}), as, for example,  in Refs. \cite{blakieprl,other3}, leads to qualitatively acceptable results for droplet and droplet-lattice formation,
but 
may lead to sizable error  in quantitative estimate of energy, size, etc.  of the final state. }


Equation (\ref{eq.GP3d}) can be reduced to 
the following  dimensionless form by scaling lengths in units of $l = \sqrt{\hbar/m\omega_z}$, time in units of $\omega_z^{-1}$,  energy in units of $\hbar\omega_z $, and density $|\psi|^2$ in units of $l^{-3}$
\begin{align}
i &\frac{\partial \psi({\bf r},t)}{\partial t} =
{\Big [}  -\frac{1}{2}\nabla^2
+\frac{1}{2}\Big(\frac{f_x^2}{f_z^2}x^2+ \frac{f_y^2}{f_z^2}y^2+z^2\Big)
\nonumber\\ &
+3a_{\mathrm{dd}}  N
\int \frac{1-3\cos^2 \theta}{|{\bf R}|^3}
\vert\psi({\mathbf r'},t)\vert^2 d{\mathbf r}'  \nonumber \\ &+ 4\pi{a} N \vert \psi({\bf r},t) \vert^2
+\gamma_{\mathrm{QF}}N^{3/2}
|\psi({\mathbf r},t)|^3
\Big] \psi({\bf r},t).
\label{GP3d}
\end{align}
Equation (\ref{GP3d})  
can also be obtained from the variational rule
\begin{align}
i \frac{\partial \psi}{\partial t} &= \frac{\delta E}{\delta \psi^*} 
\end{align}
with the following energy functional (energy per atom)
\begin{align}
E &= \int d{\bf r} \Big[ \frac{|\nabla\psi({\bf r})|^2}{2} +\frac{1}{2}\Big(\frac{f_x^2}{f_z^2}x^2+ \frac{f_y^2}{f_z^2}y^2+z^2\Big)|\psi({\bf r})|^2\nonumber
 \\
&+ \frac{3}{2}a_{\mathrm{dd}}N|\psi({\bf r})|^2 
\left.  \int \frac{1-3\cos^2\theta}{R^3}|\psi({\bf r'})|^2 d {\bf r'} \right. \nonumber \\
& + 2\pi Na |\psi({\bf r})|^4 +\frac{2\gamma_{\mathrm{QF}}}{5} N^{3/2}
|\psi({\bf r})|^5\Big]
\end{align}
for a stationary state.

\section{Numerical Results}

\label{III}

We  solve   partial differential
 equation (\ref{GP3d}) for a dipolar BEC  numerically, using FORTRAN/C programs \cite{dip} or their open-multiprocessing versions \cite{omp}, 
by the split-time-step Crank-Nicolson
method \cite{crank} employing the imaginary-time propagation rule. Often, the intensity of the system has large extension in the $x-y$ plane and it is appropriate to take a larger number of discretization steps along $x$ and $y$ directions as compared to the same along the $z$ direction.  It is problematic to treat numerically the nonlocal dipolar interaction integral in the beyond-mean-free model (\ref{GP3d}) in configuration space due to the $1/|{\bf R}|^3$ term. To circumvent the problem, this term is evaluated in the momentum space by a Fourier transformation using a convolution identity \cite{dip}, which is advantageous numerically due to the smooth behavior of this term in momentum space. The Fourier transformation of the dipolar potential in 3D can be found analytically enhancing the accuracy of the numerical procedure.


Instead of presenting  results  in dimensionless
units, we prefer to 
relate our results to the recent experimental  \cite{2d2} and related theoretical \cite{2d4} studies   on dipolar droplets   using $^{164}$Dy atoms.  For the appearance of droplets we need a strongly dipolar atom with $\varepsilon_{\mathrm{dd}}>1$  necessarily \cite{2d3}. In this study we take $a=85a_0$, close to its experimental estimate $a=(92\pm 8)a_0$  \cite{scatmes},
and $a_{\mathrm{dd}}=130.8a_0$, where $a_0$  is the Bohr radius; consequently, $\varepsilon_{\mathrm{dd}}=1.5388... >1$.
This value of scattering length is close to the scattering lengths   $a=88a_0$ \cite{2d2,2d4} and $a=70a_0$ \cite{blakieprl}
used in some other  studies of quantum droplets in a quasi-2D dipolar BEC. 
The trap frequencies along $x,z,$ and $y$ directions  are 
taken as $f_x=33$ Hz, $f_z=167$ Hz, and $f_y= 110$ Hz (trap $U_A$),  = 60 Hz (trap $U_B$), = 33 Hz (trap $U_C$)  as in recent experimental \cite{2d2} and theoretical \cite{2d4} investigations on   triangular-lattice crystallization of droplets. 
The trap $U_A$ is of  quasi-1D type along the $x$ direction  ($f_y,f_z\gg f_x$) and the trap $U_C$ it is of a cylindrically-symmetric quasi-2D type in the $x-y$ plane ($f_x=f_y \ll f_z$).  The trap $U_B$ is  an asymmetric  trap  ($f_x\ne  f_y   \ne f_z$) in the transition domain from a quasi-1D to a quasi-2D type.  With these parameters $-$ frequencies for trap $U_C$  and scattering length $a$ $-$ we found simultaneously  square-, triangular-, and honeycomb-lattice crystallization of droplets in a trapped quasi-2D dipolar BEC of $^{164}$Dy atoms  
and these three different arrangements of droplets are found to have  similar energies for a fixed $N$. 
 In this study we have  $m(^{164}$Dy)    $=164 \times 1.66054\times 10^{-27}$ kg, $\hbar =  1.0545718 \times  10^{-34}$ $m^2$ kg/s, $\omega_z = 2\pi \times 167 $  Hz, consequently, unit of length $l=\sqrt{\hbar/m\omega_z}= 0.607$  $\mu$m.

For an efficient and quick convergence of a single-droplet state or of a lattice-droplet arrangement  in an imaginary-time calculation,  an appropriate choice of the initial state is essential. The numerical simulation of a single-droplet state was started with a Gaussian wave function of small width:
$\phi({\bf r}) \sim e^{-x^2-y^2}e^{-z^2/\alpha^2},$
with the width parameter $\alpha\approx 4$. 
The numerical simulation for 
 a lattice-droplet state was started by many  Gaussian droplets  arranged on a desired lattice. 
For example,   a 49-droplet square-lattice state, viz. Fig. \ref{fig3}(f),   was started with the following analytic function
\begin{align} \label{ana}
\phi({\bf r})\sim \sum_{i,j=0}^{ \pm 1,\pm 2,\pm 3}   e^{-(x+\beta i)^2-(y+\beta j)^2}e^{-z^2/\alpha^2}, 
\end{align}
with the lattice length $\beta  \approx 5$. 
The calculation with a honeycomb- and triangular-lattice states were initiated similarly using analytic initial functions  with the droplets arranged appropriately.

\begin{figure}[t!]
\begin{center}
\includegraphics[width=\linewidth]{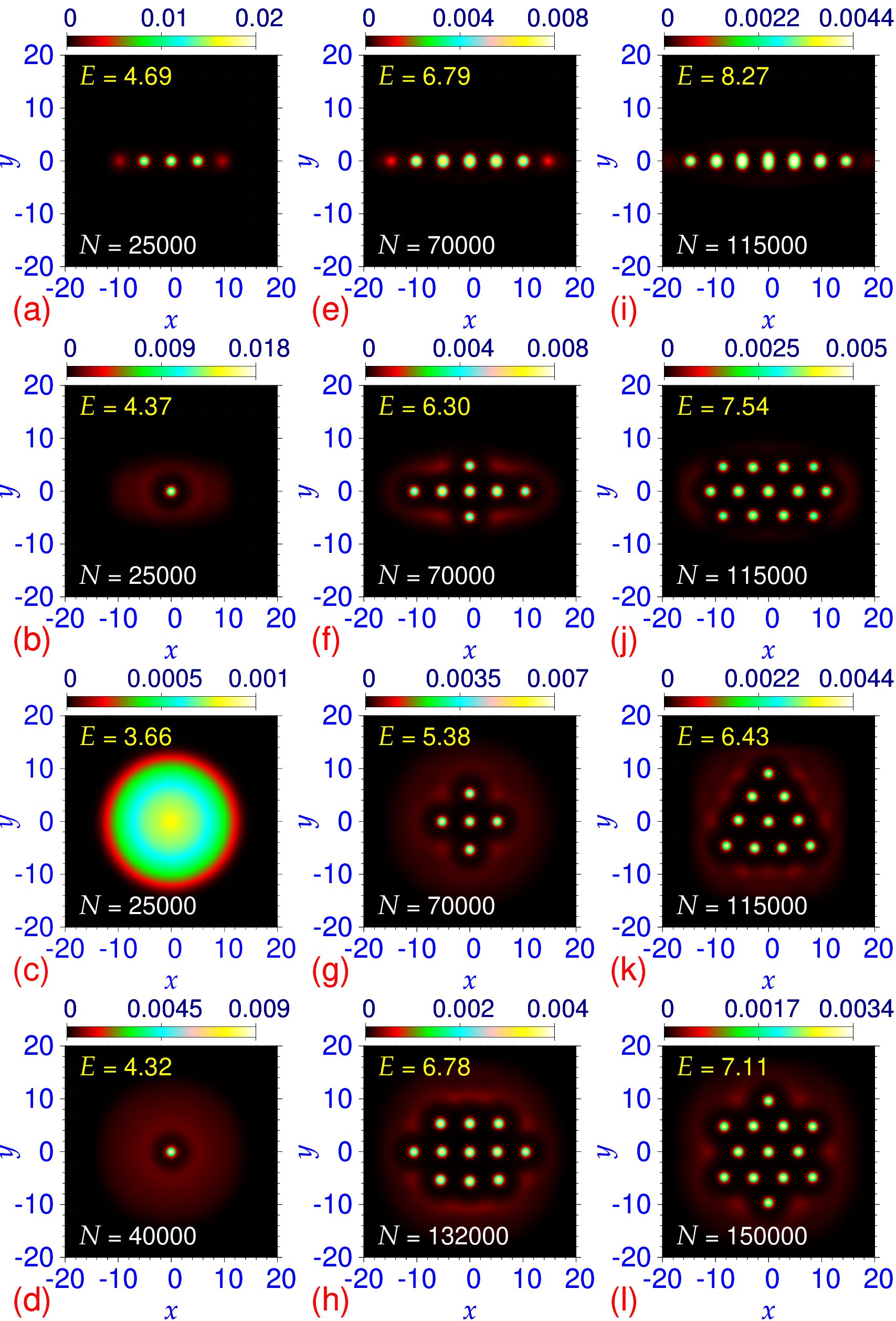}

\caption{ Contour plot of density $|\psi(x,y,0)|^2$ of a dipolar BEC of $N=25000$ $^{164}$Dy atoms  in trap  (a)  $U_A (f_y=110$ Hz), (b)  $U_B (f_y=60$ Hz) and 
(c) $U_C (f_y=33$ Hz); the same of  $N=40000$ atoms in trap (d) $U_C$; the same of $N=70000$ atoms in  trap  (e)  $U_A$, (f)  $U_B$ and 
(g) $U_C$;  the same of  $N=132000$ atoms in trap  (h) $U_C$; 
the same of $N=115000$ atoms in  trap  (i)  $U_A$, (j)  $U_B$ and 
(k) $U_C$;  the same of $N=150000$ atoms in trap (l) $U_C$. All plots are labelled by  respective  $E$ and $N$ values.
 In the first column we  illustrate the formation of a single droplet, in the second column the formation of square-lattice arrangement of droplets and in the third column that of triangular-lattice arrangement of droplets. Other parameters in all calculations are  $f_x=33$ Hz, $f_z=167$ Hz, $a=85a_0, a_{\mathrm{dd}}=130.8a_0$.
{ Plotted quantities in all figures  are dimensionless}; the length scale $l\equiv\sqrt{\hbar/m\omega_z}=0.607$ $\mu$m.
}
\label{fig1} 
\end{center}
\end{figure}

To find an 1D crystallization of droplets, we consider  25000 $^{164}$Dy atoms in 
 the quasi-1D trap  $U_A$.  With this trap $f_x \ll  f_y, f_z$,  the dipolar BEC crystallizes in droplets along the $x$ axis.  The  converged final state in this case can be obtained by imaginary-time simulation using an initial  Gaussian wave function.  However, the convergence is quicker if we use an 
analytic wave function for  a few droplets (3 or 5) periodically arranged  along the $x$ direction with a mutual 
separation $\beta$ and symmetrically placed around the occupied $x=0$ site.   A contour plot of the $z=0$ section of the 3D density 
$|\psi(x,y,0)|^2$ is shown in Fig. \ref{fig1}(a) with 3 droplets placed symmetrically around $x=0$ (a parity-symmetric state).
 For the  same set of parameters, there is a parity-antisymmetric excited state of higher energy with four droplets placed symmetrically around $x=0$,  
but   with the central site at $x=0$ unoccupied (not shown here), viz. Fig. 1(a) of Ref. \cite{2d4}.  
 In trap $U_B$, as  trap frequency  $f_y$ is reduced to $60$ Hz, the number of droplets for $N=25000$ reduces from 3 to 1 as shown in Fig. \ref{fig1}(b), where we use the final converged wave function of Fig. \ref{fig1}(a)   as the initial state in the imaginary-time simulation.
A droplet will be formed when the density is larger than a critical density. 
Inside a droplet the dipolar interaction is so strong that the dipolar BEC becomes quasi-1D along the $z$ direction  with a  small transverse section. 
A weaker trap in Fig. \ref{fig1}(b), compared to the same in Fig. \ref{fig1}(a), requires   a larger  number of atoms to attain the critical density 
required for  droplet  formation \cite{other1}.  Consequently, a droplet in the weaker trap $U_B$ can accommodate a larger number of atoms and the number of droplets is reduced from 3 in trap $U_A$ to 1
 in trap $U_B$. In  the cylindrically-sympra-clean.tex metric quasi-2D weak trap $U_C$,  the central density for 25000 atoms
is smaller than the threshold for droplet formation; consequently,   
 no  droplets can be formed and the density is of normal Gaussian type, viz. Fig. \ref{fig1}(c),  with a large increase in the size of the condensate. However, for {$N>N_{\mathrm{cr}}=33000$}  the critical density for the formation of droplet is attained in trap $U_C$ and a droplet can be formed as shown in Fig. \ref{fig1}(d) for $N=40000$.

{
We illustrate the quasi-1D to quasi-2D transition of square-lattice arrangement of droplets for a fixed number of atoms $N=70000$ in Figs. \ref{fig1}(e)-(g) for traps $U_A$, $U_B$ and $U_C$, respectively, through a contour plot of 2D density $|\psi(x,y,0)|^2$ in the $x-y$ plane.  In the quasi-1D trap $U_A$, we have a linear chain of droplets in Fig. \ref{fig1}(e) and, in the quasi-2D trap $U_C$, an $x-y$ symmetric arrangement of droplets is obtained as shown in Fig. \ref{fig1}(g).  An arrangement of droplets in the quasi-1D to quasi-2D transition domain in trap $U_B$ is illustrated in Fig. \ref{fig1}(f). The number of droplets in a specific trap  increases with $N$ as shown in Fig. \ref{fig1}(h) for $N=132000$ in trap $U_C$ with 11 droplets compared to 5 droplets in Fig. \ref{fig1}(g) for $N=70000$. 

 }

\begin{figure}[t!]
\begin{center}
\includegraphics[width=\linewidth]{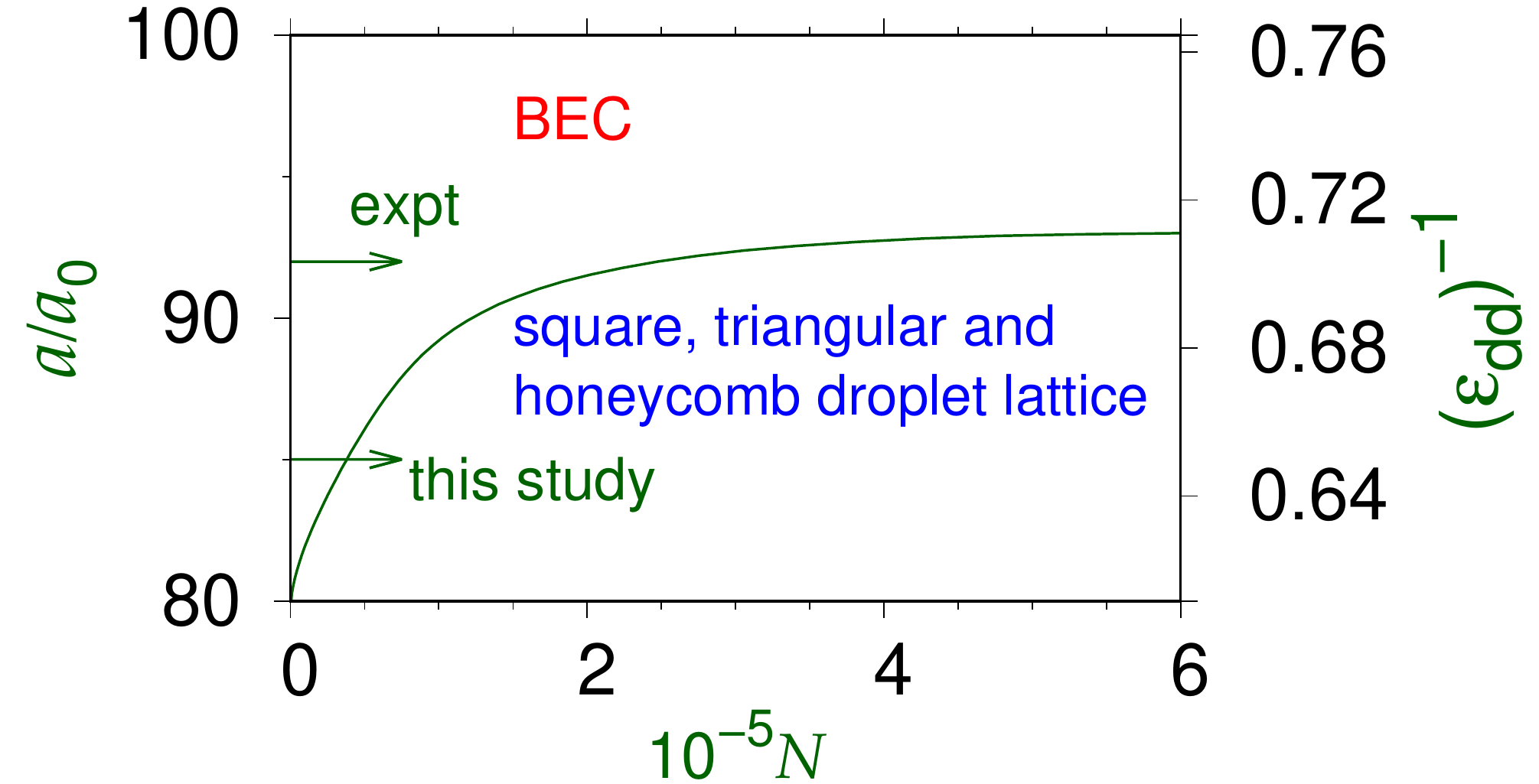}
\caption{The $a/a_0$ versus $N$ phase plot of droplet formation in trap $U_C$ illustrating the square-, triangular- and honeycomb-lattice  states. The region marked 
``BEC'' represents a normal  BEC of the type displayed in Fig. \ref{fig1}(c), where no droplet can be formed.
{ }}
\label{fig2} 
\end{center}
\end{figure}

{
The quasi-1D to quasi-2D transition of triangular-lattice arrangement of droplets for  $N=115000$ in Figs. \ref{fig1}(i)-(k) for traps $U_A$, $U_B$ and $U_C$, respectively, is considered next through a contour plot of 2D density $|\psi(x,y,0)|^2$ in the $x-y$ plane. In the quasi-1D trap $U_A$ again we have a linear array of droplets in Fig.  \ref{fig1}(i) and a triangular lattice of droplets in the quasi-2D trap $U_C$ is displayed in Fig. \ref{fig1}(k).   An intermediate triangular-lattice arrangement of droplets in trap $U_B$ in the transition  from quasi-1D to quasi-2D is presented in Fig. \ref{fig1}(j). 
In a quasi-1D trap $U_A$, the number of droplets $n_d$ increases with  $N$, as can be found from Figs. \ref{fig1}(a), (e) and (i). In the quasi-2D trap $U_C$, in general $n_d$ also increases with $N$,  viz.  Figs. \ref{fig1}(g), (k), (h), and (l) with 5, 10, 11, 13 droplets for $N=70000, 115000, 132000, 150000$, respectively. {The number of droplets is roughly proportional to the number of atoms.} 
 In Figs. \ref{fig1}(k) and (l) we find that a triangle-shaped triangular lattice has changed to a star-shaped triangular lattice with the increase of $N$ in the same trap $U_C$. 

In all cases, surrounding the lattice arrangement of droplets,  a cloud of atoms is found.  It was not possible to avoid this cloud;  if the calculation is repeated with a smaller $N$, the droplets at the four corners start to disappear maintaining the cloud intact.  Similar cloud was also found in other theoretical  \cite{2d4,other1,blakieprl} and experimental \cite{2d2} investigations.  

}

\begin{figure}[t!]
\begin{center}
\includegraphics[width=\linewidth]{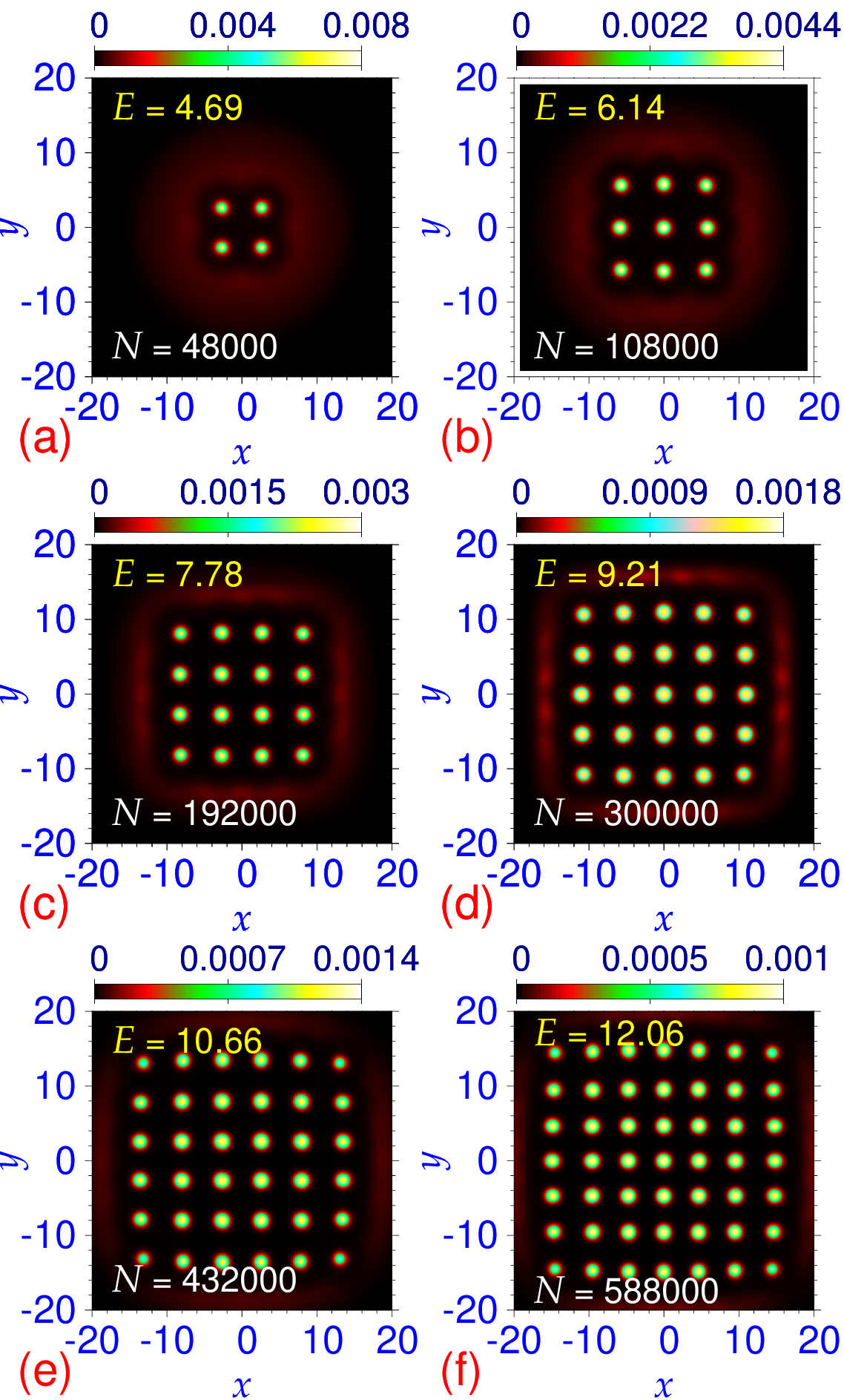}

\caption{ Contour plot of  density $|\psi(x,y,0)|^2$ of square-lattice crystallization of 4, 9, 16, 25, 36, and 49 droplets, respectively,  
for (a)  $N=48000$, (b)   $N=108000$,
(c)  $N=192000$, (d)   $N= 300000$, 
   (e)  $N=  432000$   (f) $N= 588000$ in trap $U_C$. 
}
\label{fig3} 
\end{center}
\end{figure}
 
 { After having established the formation of dipolar droplets of different symmetries, we present a $a/a_0$ versus $N$ phase plot 
 for  droplet formation  in trap $U_C$, (as our study will be confined to this quasi-2D trap,)    in Fig. \ref{fig2}, where the experimental scattering length $a=92a_0$  \cite{scatmes} and the present scattering length $a=85a_0$  are marked by arrows.  This phase plot bears some similarity  with the  phase plot
presented in { Fig. 3 of} Ref. \cite{other1} for $a_{\mathrm{dd}}=130a_0$ {in spite of different trap frequencies in  that reference}.  Although the region of droplet formation of Ref. \cite{other1} is quite similar to the same of Fig. \ref{fig2}, only triangular-lattice formation is reported in Ref. \cite{other1}. Here  we show that  it is possible to have periodic square-, honeycomb-, and triangular-lattice arrangement of droplets in the same  region. The stripe and honeycomb structures (and not honeycomb-lattice droplet as reported in this paper)
of Fig. 3 of Ref. \cite{other1} are possible beyond $N=6\times 10^5 $, which is not considered in Fig. \ref{fig2}.}  
 In this study, we employ   $a=85a_0$  and a large $N,$ deep inside the region of droplet formation in Fig. \ref{fig2}, where a large number of droplets can be formed. 

\begin{figure}[t!]
\begin{center}
\includegraphics[width=\linewidth]{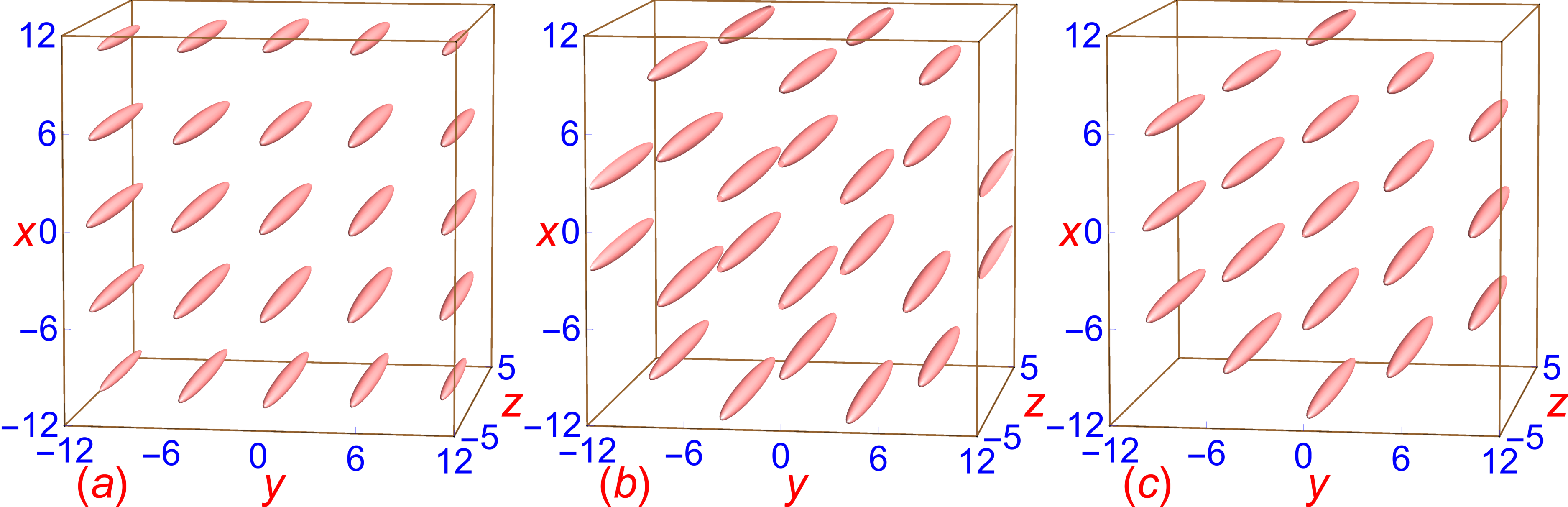}

\caption{ Isodensity contour of 3D density $|\psi(x,y,z)|^2$ for (a)  the square-lattice crystallization of Fig. \ref{fig3}(d) for $N=300000$,  (b)  the honeycomb-lattice crystallization of Fig. \ref{fig5}(b) for $N=264000$ and (c) the triangular-lattice crystallization of Fig. \ref{fig6}(b) for $N=228000$ in trap $U_C$.  
}

\label{fig4} 
\end{center}
\end{figure}

To study the square-lattice crystallization of droplets in the cylindrically-symmetric quasi-2D trap $U_C$ we note that there are two types of square-lattice crystallization $-$ an even number of droplets on each side of the square (with $2\times 2=4,$ $4\times 4=16,$ $ 6\times 6=36$ etc. droplets)  or an odd number of droplets on each side of the square  (with $3\times 3=9,$ $5\times 5=25,$ $7\times 7=49$ etc. droplets);$-$  the corresponding density $|\psi(x,y,0)|^2$ in trap $U_C$ is plotted in  Figs. \ref{fig3}(a)-(f) for $N= 48000, 108000, 192000, 300000, 432000,$ and 588000, respectively. 
For a symmetric distribution of the droplets,
in the first  (second) type, 
the central site  at $x=y=0$ has to be  vacant (occupied), corresponding to a parity-antisymmetric (parity-symmetric) state. All states are obtained by  imaginary-time simulation using   an analytic initial wave function on a square lattice with lattice spacing $\beta$,
 viz. Eq. (\ref{ana}).      The appropriate  $N$  per droplet in a calculation  for an efficient square-lattice formation was  found to be of the order of 12000. For a smaller $N,$ the droplets at the corners may  disappear  and for a larger $N,$ a intense cloud is formed around the droplets. With further increase of number of atoms, multiple (about two to four) droplets will be formed, thus reducing the cloud.  
In Fig. \ref{fig4}(a) we display the isodensity  contour of density $|\psi(x,y,z)|^2$, for  the square-lattice crystallization of Fig. \ref{fig3}(d).


\begin{figure}[t!]
\begin{center}
\includegraphics[width=\linewidth]{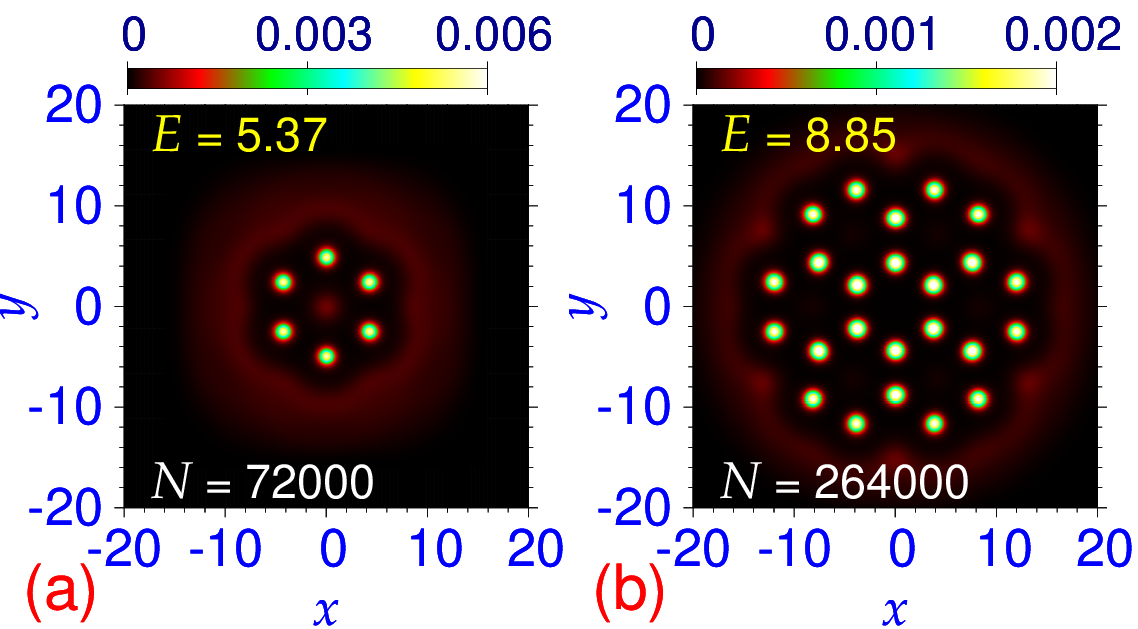}

\caption{ Contour plot of density $|\psi(x,y,0)|^2$ of honeycomb-lattice crystallization of droplets for  (a)  $N=72000$, (b)   $N= 264000$ in trap $U_C$.
}
\label{fig5} 
\end{center}
\end{figure}

\begin{figure}[t!]
\begin{center}
\includegraphics[width=\linewidth]{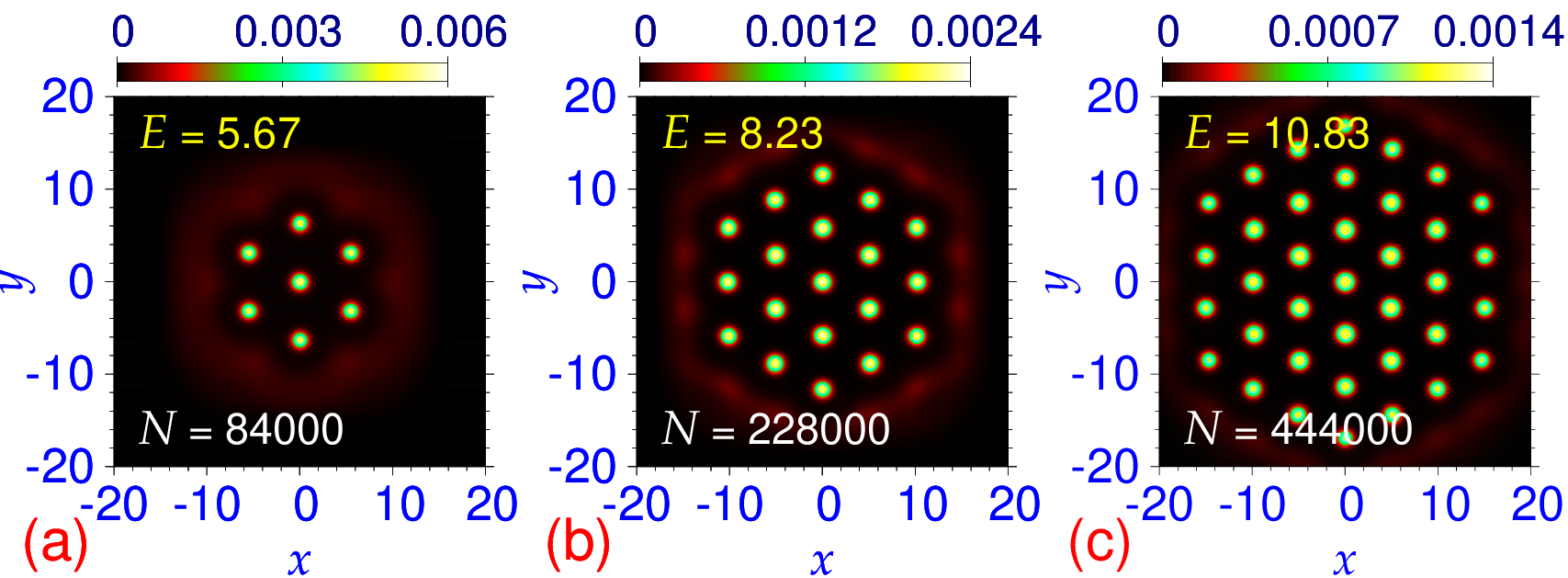}

\caption{ Contour plot of density  $|\psi(x,y,0)|^2$ of triangular-lattice crystallization of droplets    for (a)  $N=84000$, (b)   $N=228000$ and (c) $N=444000$ in trap $U_C$.
}
\label{fig6} 
\end{center}
\end{figure}

A numerical simulation of the honeycomb-lattice crystallization of  droplets   needed much more care than the square- and triangular-lattice arrangement of droplets.  This is because this arrangement is basically a triangular-lattice arrangement 
of droplets with  a missing droplet at the center of all closed adjacent hexagons    and if the initial state is not properly chosen, the imaginary-time numerical simulation may converge to the   triangular-lattice arrangement of droplets  filling in the vacant positions at the center of the closed hexagons with droplets.
    The densities of  honeycomb-lattice crystallization  for 6 and 24 droplets in trap $U_C$ are displayed  in Figs. \ref{fig5}(a) and (b), respectively, for $N=72000$ and 264000. In Fig. \ref{fig4}(b) we present  the  isodensity  contour of 3D density pra-clean.tex 
$|\psi(x,y,z)|^2$ for  the honeycomb-lattice crystallization of Fig. \ref{fig5}(b).

Finally, we investigate the   triangular-lattice crystallization of droplets  in trap $U_C$. 
In Figs. \ref{fig6}(a), (b) and (c) we display the contour plot of density $|\psi(x,y,0)|^2$ of
 triangular-lattice crystallization of 7, 19, and 37 droplets  in  trap $U_C$  for  $N=84000, 228000$ and 444000, respectively, calculated using an initial state of similar symmetry properties.  In Fig. \ref{fig4}(c) we display the isodensity   contour of density 
$|\psi(x,y,z)|^2$, for  the triangular-lattice crystallization of Fig. \ref{fig6}(b).



\begin{figure}[t!]
\begin{center}
\includegraphics[width=\linewidth]{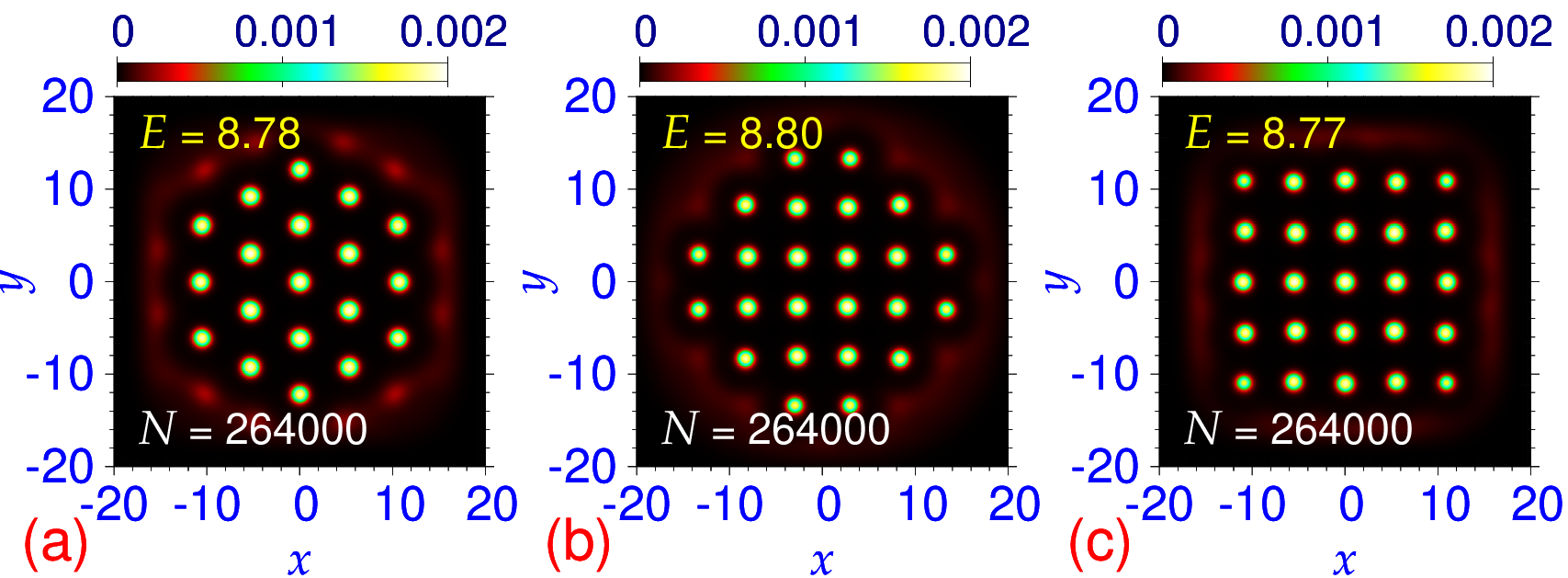}
 
\caption{ Contour plot of density  $|\psi(x,y,0)|^2$ of (a) triangular-lattice crystallization, (b) square-lattice crystallization with vacant central site, and (c)  square-lattice crystallization with occupied central site,    for   $N=264000$,  in trap $U_C$.
}
\label{fig7} 
\end{center}
\end{figure}

\begin{figure*}[htbp]
\begin{center}
\includegraphics[width=\textwidth]{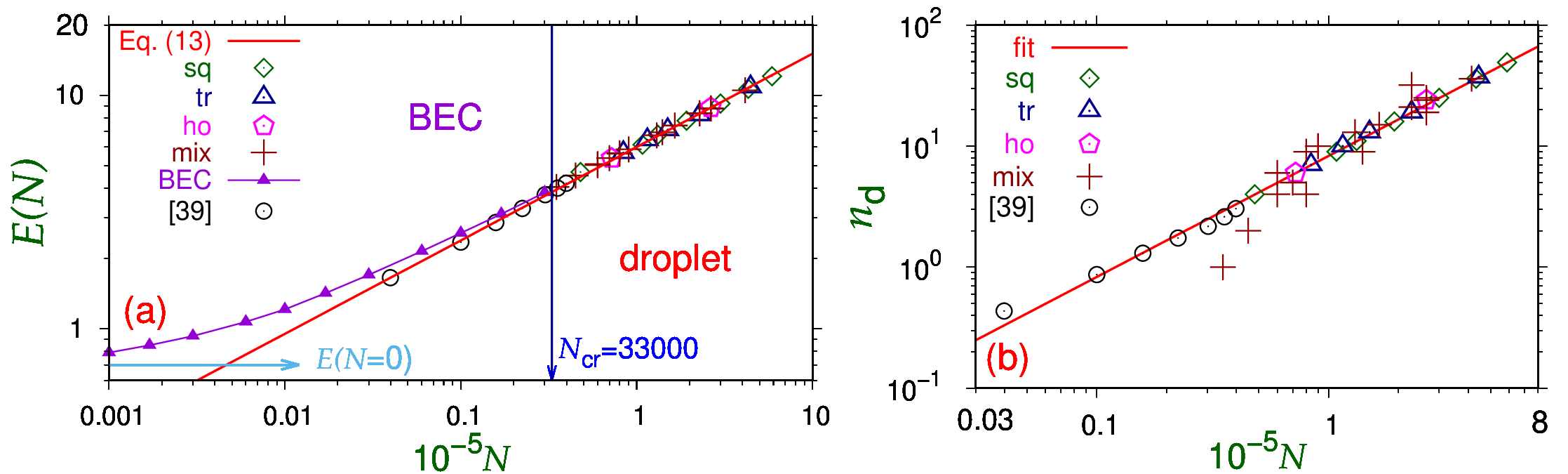}

\caption{(a) Energy per atom $E$ and (b) number of droplets $n_{\mathrm d}$ versus  $N$  of all lattice states in trap $U_C$  on log-log scale;  sq:  square-lattice states of Figs.  \ref{fig1}(g)-(h) and \ref{fig3};  tr: triangular-lattice states of Figs. \ref{fig1}(k)-(l) and \ref{fig6};   ho: honeycomb-lattice states of Fig.  \ref{fig5};  mix: the states presented in Fig.   \ref{fig7} and a few more not shown in this paper; BEC: normal BEC superfluid for smaller $N$ (states of type presented in Fig. \ref{fig1}(c)); \cite{blakieprl}: one- to seven-droplet states from Fig. 2(a) of \cite{blakieprl} employing $f_x=f_y= 60$ Hz, $f_z=300$ Hz,  $a=70a_0$, and approximation (\ref{app})   in arbitrary units. 
The points are numerical results and the straight lines labeled  (a) ``Eq. (\ref{scaling})'' and (b) ``fit''  are the scaling relations $E \approx 0.06\times  N^{0.4}$ and $n_{\mathrm{d}}\approx N/12000$, respectively. 
The regions of normal BEC and droplet formation in trap $U_C$ are marked ``BEC'' and ``droplet'', respectively.}

\label{fig8} 
\end{center}
\end{figure*}

{It is pertinent to ask that,  of the  states of  different arrangements of dipolar droplets, which is/are the ground state(s) and which are the excited states. There are two types of arrangement of droplets: the one with the central site at $x=y=0$ occupied (square lattice with an odd number of droplets and triangular lattice)  and one with the central site vacant (square lattice with an even number of droplets and honeycomb lattice).   We find that the first of  these types  forms quasi-degenerate ground states and the second type forms low-lying excited states.  To demonstrate this claim we display in Figs. \ref{fig7}(a)-(c) contour plot of density $|\psi(x,y,0)|^2$ for $N=264000$  for triangular-lattice symmetry  and two types of square-lattice symmetries. We find from Figs. \ref{fig6}(b) and 
\ref{fig7}(a) that one can have a 19-droplet triangular-lattice state for $N=228000$ and $N=264000$, respectively; 
with an increased $N$ in Fig. \ref{fig7}(a)  one has an increased density of atom cloud. Similarly, in Figs.  \ref{fig3}(d)  and \ref{fig7}(c)  one finds a 25-droplet square-lattice state for $N=300000$ and $N=264000$,
respectively. In this case with larger $N$ in Fig.  \ref{fig3}(d)  one has an increased density of atom cloud.
From Figs. \ref{fig7} and \ref{fig5}(b), for a fixed  $N=264000$, we find that the parity-symmetric square-lattice state with occupied central site, viz. Fig. \ref{fig7}(c), and  the triangular-lattice state, viz. Fig. \ref{fig7}(a),  constitute the quasi-degenerate  ground states of energy $E=8.77$ and $E=8.78$, respectively. 
The parity-antisymmetric square-lattice state with vacant central site $(E=8.80)$, viz. Fig. \ref{fig7}(b),  and the honeycomb-lattice state $(E=8.85)$, viz. Fig. \ref{fig5}(b), are low-lying excited states. {Similar result was found to be true for a few other $N$ values (details not reported in this paper).}
}{ In addition to the states of droplet arrangements on periodic lattices, highlighted in this paper, there could also be states of droplet arrangements with no specific symmetry. For a specific trap  and for a fixed  $N$ all these states have nearby energies (not illustrated in this paper). In addition to the quasi-degenerate stable ground states,  the imaginary-time approach also finds  excited states with a specific symmetry, e.g.,
the parity-antisymmetric square-lattice and honeycomb-lattice  states, which could be metastable. 

}
 

To study the universal nature of the formation of droplets for different  $N$ and different lattice symmetries    in trap $U_C$,
we plot  in Fig. \ref{fig8}(a)-(b) the energy $E$ per atom  and  the number of droplets $n_{\mathrm d}$  versus the corresponding  $N$. { In addition,  we plot the energy from Fig. 2(a) of Ref. \cite{blakieprl} for one- to seven-droplet states, in arbitrary units, calculated with different trap parameters
($f_x=f_y= 60 $ Hz and $f_z= 300$ Hz) and a different scattering length $a=70a_0$.
We could {reproduce} the results of Fig. 2(a) of  Ref. \cite{blakieprl} using the approximate auxiliary function (\ref{app}) in place of the exact expression (\ref{exa}) used in this paper. 
 For { example, { using Eq. (\ref{app}),}  for} the one-droplet state  of Ref. \cite{blakieprl}   we obtain the energy per atom $E/h\approx 348$ Hz  and for the seven-droplet state we obtain $E/h\approx 840$ Hz,  close to the results illustrated in Fig. 2(a) of  Ref. \cite{blakieprl}.} {The use of Eq. (\ref{exa}) leads to much larger  energies.}
 From Fig. \ref{fig8}(a) 
 we find the scaling relation between    $E$ and  $N$  in the region of droplet formation  {($N>N_{\mathrm{cr}}=33000$} in trap $U_C$): 
\begin{align} \label{scaling}
E \approx 0.06\times  N^{0.4}
\end{align}
 independent of the lattice symmetry, scattering length $a$, and trap parameters. 
{We multiplied the results of Ref. \cite{blakieprl} by an arbitrary factor ($\sim 1.5$) to take care of the prefactor in scaling (\ref{scaling}), nevertheless, it is remarkable that all points lie on the same universal line and  the  exponent (0.4) is independent of scattering length and trap parameters. The point with the smallest number of atoms from our calculation in Fig. \ref{fig8}(a) is $N=35000$ with one droplet. By including the results of Ref. \cite{blakieprl}, we could include six points with $N< 40000$ containing $1-7$ droplets covering about one order of magnitude in Fig. \ref{fig8}(a):  with the parameters of    Ref. \cite{blakieprl}  one droplet can be generated with only a much smaller number $N=3980$ of atoms.  }
{ The points labelled BEC in Fig. \ref{fig8}(a) represent a normal superfluid BEC without droplet formation, viz. Fig. \ref{fig2}. These points deviate a bit from the universal scaling (\ref{scaling}) valid for droplets, specially  for small $N$.}
From Fig. \ref{fig8}(b), with exactly the same points of Fig. \ref{fig8}(a), 
the number of droplets $n_{\mathrm{d}}$ in an arrangement is approximately 
linearly proportional to  $N$:  $n_{\mathrm{d}}\approx N/12000$ indicating the average number of 12000 atoms per droplet; we note that the points  generating the   large width of the scaling  in Fig. \ref{fig8}(b) have collapsed  on  the straight-line  fit (\ref{scaling}) in Fig. \ref{fig8}(a).
{ The small difference in energy between the ground and the excited states, viz. Figs. \ref{fig7} and \ref{fig5}(b), is not noticeable in Fig. \ref{fig8}(a).  We have established the universal nature of scaling (\ref{scaling}) in the fact that the  exponent is independent of not only the value of $N$ extended over about  two orders of magnitude but also of different parameters of the problem, trap frequencies and scattering length. It remains to be seen if this exponent is independent of large variation of the dipole moment or of the details of the beyond-mean-field correction, that stops the collapse.  Only after establishing the true universality,   the physical origin of the scaling relation could be addressed \cite{scaling}, which will be an interesting topic of future investigation.
   }

{
{In this  theoretical study 
we have neglected the effect of three-body recombination loss of atoms. 
A matter of concern for the experimental observation \cite{drop1,other3} of a spatially periodic lattice of droplets is the large atom number ($N\sim 10^5$) required, where the effect of  three-body recombination loss of atoms might not be negligible \cite{other1}.  Nevertheless, a reasonably small value of the loss parameter
$ (=1.25 \times 10^{-41})$ m$^6$/s is estimated for $^{164}$Dy atoms \cite{1d1,drop1} from measurements
on a thermal cloud and is assumed to be a constant over the
small range of scattering lengths   near $a=60-80a_0$ close to the experimental estimate $a=92a_0$ \cite{scatmes} and the value $a=85a_0$ used in this study. Considering an upper estimate of  $N=10^6$ atoms
of this study, viz. Figs. \ref{fig3} and \ref{fig4},
in a volume of $40\times 40\times 10$ in dimensionless units with the length scale $l=0.607$ $\mu$m, we obtain a typical  atomic density of  $3\times 10^{14}$ cm$^{-3}=3\times 10^{20}$ m$^{-3}$ within the acceptable limit for the formation of droplets in  an experiment
as established  in previous experimental \cite{drop1,1d1} and theoretical \cite{other1,other3} investigations.}
}

\section{Summary}

\label{IV}

We have demonstrated, using the  GP equation including the  quantum-fluctuation LHY interaction,   supersolid-like spatially-periodic  crystallization of droplets of a cylindrically-symmetric quasi-2D trapped dipolar BEC on square and honeycomb lattices, in addition to the triangular-lattice crystallization observed experimentally \cite{2d3} and studied theoretically \cite{2d4,blakieprl}. There are two possible types of square-lattice crystallization of droplets, e.g., with the central site {at $x=y=0$} occupied (parity-symmetric) or vacant (parity-antisymmetric). The parity-symmetric square-lattice crystallization and triangular-lattice crystallization form two quasi-degenerate ground states.  The parity-antisymmetric square-lattice crystallization and honeycomb-lattice crystallization, both with the central site vacant, form two low-lying excited states.   The number of droplets in these lattice arrangements increases with the  number  of atoms in an approximate linear fashion. 
We  establish a robust  scaling relation (\ref{scaling}) valid for about two orders of magnitude  between the energy per atom and the number of atoms   in the region of droplet formation, independent of the  lattice symmetry (square, honeycomb or triangular) of droplets, so that the three lattice crystallizations  should be of phenomenological interest.  
{The stability of each of these crystallizations can be theoretically established by a linear stability analysis. However, this is a formidable task of future interest,  considering the nonlocal nature of dipolar interaction.  Nevertheless, both the triangular- and square-lattice structures  are close-packed quasi-degenerate structures  with a predominantly repulsive   ``inter-droplet'' interaction and are expected to be stable, whereas, the excited honeycomb-lattice structure  has an empty site at the center of an  hexagon and is conjectured to be unstable.}
Hence,  from an energetic consideration, the {parity-symmetric} square-lattice crystallization, { with the central site at $x=y=0$  occupied,}
 is a  likely candidate for experimental observation in addition to  the already observed triangular-lattice crystallization.
 The results of this paper can be tested in experiments with strongly dipolar atomic  BECs of $^{164}$Dy or $^{168}$Er atoms
 with present knowhow.

\begin{acknowledgments}
SKA
 acknowledges support by the CNPq (Brazil) grant 301324/2019-0, and by the ICTP-SAIFR-FAPESP (Brazil) grant 2016/01343-7

\end{acknowledgments}



\end{document}